# Procedural Music Generation Systems in Games

Shangxuan Luo[1], Joshua Reiss[1]

[1]Queen Mary University of London

Correspondence should be addressed to Shangxuan Luo (`s.luo@qmul.ac.uk`)

**ABSTRACT**

Procedural Music Generation (PMG) is an emerging field that algorithmically creates music content for video games. By leveraging techniques from simple rule-based approaches to advanced machine learning algorithms, PMG has the potential to significantly improve development efficiency, provide richer musical experiences, and enhance player immersion. However, academic prototypes often diverge from applications due to differences in priorities such as novelty, reliability, and allocated resources. This paper bridges the gap between research and applications by presenting a systematic overview of current PMG techniques in both fields, offering a two-aspect taxonomy. Through a comparative analysis, this study identifies key research challenges in algorithm implementation, music quality and game integration. Finally, the paper outlines future research directions, emphasising task-oriented and context-aware design, more comprehensive quality evaluation methods, and improved research tool integration to provide actionable insights for developers, composers, and researchers seeking to advance PMG in game contexts.

## 1 Introduction

Procedural Music Generation (PMG) originates from the concept of Procedural Content Generation (PCG), which is the algorithmic creation of game content with limited or indirect user input [1]. The basic idea of PCG seems to first appear concurrently with the early development of computers around the 1980s [2]. One early example of using PCG was Rogue in 1978, which became the beginning of Rogue-like games. In Rogue, algorithms were used to generate randomised dungeons, with the map changing each time you played. Without storing massive static assets, PCG addressed hardware memory bottlenecks in the early days, but has since become more widely used to expand the game content and richness, including levels, characters, stories and sound.

In addition to significantly impacting the player's immersion by generating diverse content, PCG also dramatically reduces the development time and cost [1], [3], [4], particularly benefits the non-linearity characteristic of games that require dynamic, replayable, or expansive content, such as Rogue-likes, Sandboxes, Open-World Games, etc.

Music, as an important aspect of game content, provides unique functional support to enhance the game experience. For example, theme music aims to express the core tone or character identity of a work, whereas background music seeks to render the atmosphere of a scene. Moreover, music offers a unique means of enhancing storytelling through its ability to adapt dynamic environmental changes in games and facilitate smoother scene or state transitions.

However, most games comprise a soundtrack of between one and four hours of music, whereas the duration of game-play can vary considerably, from dozens to hundreds of hours [5]. Furthermore, Live Service Games are often designed to provide endless player experience by iterating the gameplay [6]. Yet very little music goes with it. This results in players hearing music repeated on many occasions. While repetition is a critical element of music, excess can disrupt immersion and become a source of frustration and fatigue for the player [7], [8]. Therefore, extending the concepts and methodologies of PCG to music is a highly effective strategy that alleviates this fatigue by generating a wide variety of unique content, thus improving the overall game quality.

This paper aims to examine the evolution of PMG systems in both research and application areas. It intends to compare current methodologies, highlight emerging trends, identify gaps and challenges, and provide constructive recommendations for future advancements in the field.

## 2  Background

### 2.1 Definition

An early definition of procedural music given by Collins is a composition that evolves in real-time according to a specific set of rules or control logics [9]. To further elucidate the evolution of the system in response to input data, this paper adopts a definition closer to procedural audio generation [10]: PMG is *the use of algorithms to create or modify musical content so that it adapts to changing inputs in real time*.

Compared to similar concepts like algorithmic music or generative music, which also involve creating music through systemic automation, the critical characteristic of PMG is its real-time adaptability to dynamic game content. This requires systems capable of making instant adjustments and variations to the music in response to changes in the game state. As a result, PMG often relies on complex rule-based frameworks in larger games to determine how to interpret and respond to game data, ensuring the music remains cohesive and contextually appropriate throughout the player's experience.

A typical workflow of PMG systems is illustrated in Figure 1. The system receives two primary types of input: in-game state data and music asset data. Based on these inputs, the system generates real-time

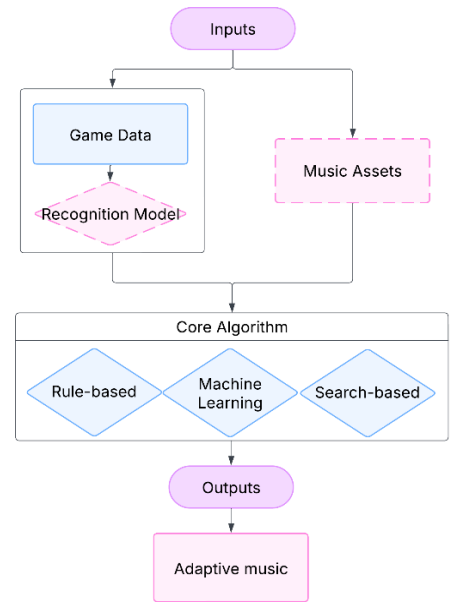

Figure 1. A Typical Workflow of PMG Systems.

adaptive music as its output, dynamically adjusting to reflect the evolving context and gameplay.

The state data usually refers to all the information used in a video game to describe the current condition of play. The following parameters are commonly used to define and control various elements of procedural music:

- *Character* parameters include vitality status (healthy or dangerous), behavioural patterns (e.g. sneaking or attacking), etc.
- *Environment* parameters include weather, time of day, terrain, etc.
- *Spatial* parameters include enemy distance, location coordinates, etc.
- *Game mechanics* parameters include quest progression, difficulty level, etc.
- *Interaction* parameters include player input (mouse/keyboard mapping) and the results of interactions with other events or objects.

In academic research, many systems introduce an intermediate model, which we tentatively call cognitive models, to interpret and analyse the above parameters to generate a reference that guides the subsequent music generation [11], [12], [13], [14]. The main references involve emotion labels and intensity levels. Essentially, this approach maps high-dimensional data into a low-dimensional space to reduce system complexity for prototyping.

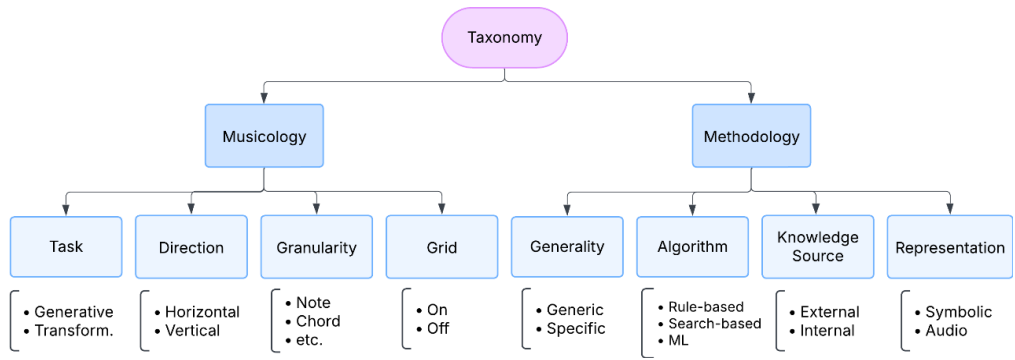

Figure 2. Hierarchy Taxonomy of Procedural Music Systems.

Another type of input data, music assets, are audio files or resources that are used in the game development process, mainly including sound effects and soundtrack clips. There are two types of musical compositions in games that indicate whether music assets are involved in the generation progress. Generative composition refers to music created entirely by algorithms without using pre-existing music assets as inputs. In contrast, transformational composition involves rearranging or modifying existing fragments from music assets to generate new music [15].

Based on the input of game data and music assets, the system dynamically generates music through its core algorithms. These algorithms are classified into three main categories: rule-based, search-based and machine learning. However, a mixed approach is commonly used in practice. We will further discuss them in the subsequent sections.

### 2.2 Taxonomy

To systematically categorise and analyse PMG systems, we adopt a comprehensive taxonomy that inherits and adapts the work of [16], shown in Figure 2. For a detailed comparison of the systems studied, see Table 1 in Appendix. We investigated eight dimensions of these systems in two main aspects: Musicology and Methodology. Musicology contains four categories, including Task, Direction, Granularity and Grid. Methodology also contains four categories: Generality, Algorithm, Knowledge Source and Representation.

*Tasks* indicate whether the system is generative or transformational [15]. Transformational systems generate new music by reorganising existing music assets, whereas generative systems usually do not require music assets as input and can generate music from scratch using rules and algorithms. The boundary between the two is not entirely deterministic, and is mainly measured by the granularity of the music manipulated by the system.

*Direction* refers to whether the system can manipulate the music horizontally or vertically. Horizontal manipulation involves generating or rearranging music over time, such as adjusting the order of measures or phrases. In contrast, vertical manipulation involves changing the music in terms of layer at a single time point, such as instrument groups.

*Granularity* refers to the level of detail with which a music system generates musical elements or aspects. A generative system has higher levels of detail in terms of note compared to a transformational system, which often manipulates the music at a coarser level, such as measures or phrases. Other aspects of granularity include chords, timbre, tempo, velocity, etc.

*Grid* refers to whether the music generated by the system is aligned to the beat. In other words, whether the music has a relatively fixed groove. In Western classical music, the rhythm is mainly triple and quadruple meter, and the music's tempo, dynamics, and chords are usually based on the meter. Non-aligned music is common in some Eastern cultures, but in the context of game music, it is generally for ambient establishment.

*Generality* refers to the purpose for which a system is designed. Specific systems are designed for a particular game or style of play, whereas generic systems are designed to provide a more general solution for a wider range of game scenarios or categories.

*Algorithms* refer to the methodology used by the system and are categorised into rule-based, search-based and machine learning. Rule-based methods refer to techniques according to predefined rules or patterns. Some stochastic methods, such as Markov models, are often integrated into rule-based methods. Search-based methods explore a vast space of possible solutions to find the best or most desirable outcome based on specific criteria or fitness functions. Machine learning methods simulate neural networks and are trained on large-scale data to learn music patterns.

*Knowledge Source (KS)* is where information of the algorithm comes from. External knowledge comes from composers or system designers, especially in rule-based systems. They predefine the rules based on their experience and understanding of the music, which is often complex and done manually. Internal knowledge comes from the data itself, for example, the hidden state of a stochastic model or the weight of a trained deep-learning model.

*Representation* refers to the method that a system uses to store and organise its musical knowledge. In computing systems, the most widely used format for symbolic representation is MIDI (Musical Instrument Digital Interface), where each musical event is encoded using variables such as pitch, velocity, channel, and on/off states. Audio representation involves capturing analog signals of real-world music and converting them into a digital format for storage and processing.

## 3   Research

In academia, a variety of advanced methods are used. We broadly classify them into three categories, rule-based, search-based, and machine learning-based, according to the research in PCG [1].

### 3.1 Rule-based methods

Rule-based methods, sometimes referred to as constrictive methods, usually act as constraints by encoding rules from music theory. One example is the use of a feasibility equation to limit the content generated [17]. This equation encodes the following rules:

$$Feasibility = -\sum_{n=0}^{N-1} L(n, n+1) + S(n, n+1) + D(n, n+1))$$

(1)

Where L encodes that it should not have leaps between notes bigger than a fifth, S encodes that it should contain at least a minimum number of leaps of a second and $D$ encodes that each note pitch should differ from the preceding note pitch. All of the sub-equations in the equation are Boolean, and thus the whole function returns a range of values from -3 to 0. A 0 means that the music generated is reasonable, and a negative value does not.

Lopez uses Max/MSP, a visual programming environment, to implement a rule-based system that generates real-time game music and provides interesting generation results [18]. The rhythmic development algorithm is based on Lerdahl & Jackendoff's Generative theory [19]. Pitch Generation utilises 'drunk ' and "drunk-contour" modes for real-time pitch assignment, with parameters for step, range, and non-repetition. The drunk mode based on the idea of random-walk generates discrete sequences of notes, where each note is independent and random. The drunken contour pattern generates continuous melodic contours, with randomly generated key points connected by interpolated notes.

Rule-based methods are strongly interpretable and extendable. Researchers can easily control and shape the output music in terms of different elements without affecting other aspects. However, they require high maintenance and are limited to the designer's musical knowledge.

### 3.2 Search-based methods

The most common search-based method, Genetic Algorithm (GA), is adopted in many game music studies [17], [20], [21], [22]. Musical generation is represented as an optimisation problem, and the search space here is the set of all possible music pieces generated.

GA can be non-deterministic by providing a diverse set of solutions. It creates a varied population of candidate music to adapt creative tasks. However, this requires delicate fitness function design considering both the quality and novelty of a music piece. Three fitness equations can be used to evaluate the individual music piece of a population in GA:

- Rule-based fitness function - Rules extracted from the original material or external music theory rules.

- Human fitness function - based on direct feedback from the composer [23].
- Learned fitness function - Training a neural network to decide [24].

An example of a rule-based fitness equation is:

$$C_{\text{Step}} = \sum_{n=0}^{N-1} I(n, n+1)(P(n, n+1) + Q(n, n+1))/L \quad (2)$$

Where $C_{Step}$ stands for Counter Step; $P$ for Pre-Counter Step; $Q$ for Post-Counter Step; and $L$ for Leap. All sub-equations are Boolean, and the equation evaluates whether the melody has a nice smooth contour, specifically, whether it reverses motion after a big jump (greater than a second interval). It is a common technique in musical progressions to reverse the movement of the melody after a big jump in order to make the musical progression aurally unobtrusive. Other rule-based fitness functions include melodic novelty, note density, etc. [21].

Unlike pure rule-based systems, rule-based fitness functions provide a more indirect way to select desired music, leaving more space for interesting music candidates. However, it involves high-latency iterations, making real-time deployment challenging.

When more than one fitness equation needs to be satisfied at the same time, the problem is programmed as a multi-objective optimisation (MOO), which can be solved by a method like Non-dominated Sorting Genetic Algorithm II (NSGA-II) [25]. This algorithm is designed to find Pareto bounds - the set of optimal solutions.

However, there are exceptions that use only crossover and mutation techniques of GA due to the subjective nature of fitness equations [26].

### 3.3 Machine learning

There has been a growing trend of using machine learning methods in game music systems, especially transformer architectures in recent years [11], [12], [27]. This architecture shows surprising potential in commercial areas on large-scale training, such as sunoAI and Udio. However, the data hunger of machine learning and the difficulty in controllability have been research hindrances.

Traditional neural-network architectures have not been completely abandoned either. [20] uses a multi-agent RNN model to handle melody, harmony and other tasks separately. [28] also used RNN to implement a simple adaptive music generation system. [11] used CNN to identify the mood of the game screen, combined with a transformer to generate music. Research using simple probabilistic models such as Markov chains for music generation dwindled.

### 3.4 Hybrid Method

When involving varied tasks in a complex system, it is common to use a hybrid approach to tackle different challenges [12], [18], [20], [29]. Normally, researchers adopt distinct methods to handle the generation and recognition tasks, for example, interpreting game status as emotion or intensity using a recognition model (Figure 1).

In [20], Hutchings proposed a novel Adaptive Music System (AMS) that combines cognitive models with a multi-agent composition system. The cognitive model is responsible for identifying the emotion of the game content, while three agents cooperate on melody, harmony and percussive rhythm tasks to adapt the pre-composed music to match the emotion in real-time. He incorporated a variant of genetic algorithms, Wilson's eXtended learning Classifier System [30], into his melody agent, while Recurrent Neural Networks were the backend in other agents.

## 4 Applications

In commercial games, most existing procedural music generation systems are rule-based. These rules are designed to be based on the composer or sound designer's musical knowledge and are game-specific. Compared to the advancement of PCG in the design of levels, terrain, characters, etc., game music is less well regarded and easier to adhere to traditional methods. However, many great games show the potential of applying PMG technology.

Spore is a successful early example of implementing PMG. The prototype was completed in Max/MSP, and adapted to a customised version of PureData [31]. In Spore, the music is stochastically generated in real time from many music samples through a set of rules [9], [31]. Leonard Paul was inspired by Spore and also used PureData to construct his procedural music system for Sim Cell [32]. Graphical music tools like Max/MSP and PureData have also been widely used for implementing PMG in other games, for example, The Legend of Zelda: Tears of the Kingdom [33].

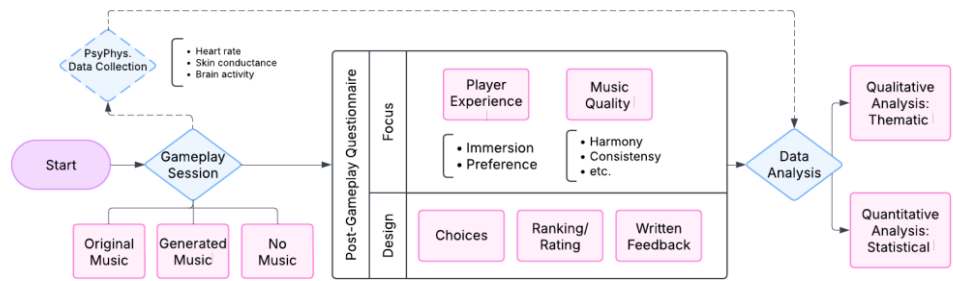

Figure 3. A Flowchart of Evaluation Design.

Proteus [34], DOOM [35] and Ape Out [36] all use timbre manipulation as a way of creating variations in the content of the music, expanding somewhat into the realm of procedural audio generation. Proteus is creative in its use of various effects to change the timbre and arrange the music in a non-aligned manner. DOOM creates sequences of varying sounds by stacking many different effects on top of each other, and then samples from a different sequence position each time using a simple rhythmic pattern to achieve a constantly changing timbre. Ape Out generates varied musical content by associating in-game action with percussion sounds.

Both Ape Out [36] and Rez Infinite [37] are highly adaptive music systems, which can also be referred to as reactive music systems, where changes in music or sound are triggered by actions/key presses. Reactive music is a special case of adaptive music, more like an interactive synthesiser, commonly found in rhythm games [16]. Here, the input is mainly the player's interaction data. Unlike Ape Out's timbre mapping, music and sound in Rez Infinite are primarily pitch mapped and are generated procedurally by responding to the player's shooting action with a constant rhythmic beat that synchronises with the character's vibration.

Genesis Noir [38] creatively takes advantage of Jazz improvisation in its music system and brings a unique solution for PMG systems. Recorded music based on preset chord progression provides the overall music ambience, while the player's reaction generates the leading melody in real time. There are two ways music adapts to the gameplay, one of which is by responding to a series of players' inputs through dialogue. For example, the player clicks several buttons to generate a melody sequence, and the system re-arranges it in terms of rhythm and decoration; another way is by instantaneously responding to the interaction. For example, the mouse's motion speed controls the rhythm, and its y-axis coordinates on the screen control the pitch.

Whilst audio middleware is a versatile solution for deploying adaptive music for small to medium-sized games, large games require more complex and flexible solutions to arrange music. As a consequence, they develop independent systems that can integrate seamlessly with the game engine and handle all the interaction logic on how the music adapts to changes in the game. For example, Music Manager in Ara: History Untold [39] is a data-driven system combining both a weighting scheme and stochastic rules. It collects data within a turn and makes a decision about music generation at the end of each turn. Other applications include Pulse in NMS [40], and a server-based music system in The Outlast Trials [41].

## 5  Evaluation Design

One crucial aspect that differentiates research from application is that researchers typically employ a rigorous assessment to evaluate the effectiveness of PMG systems, which involves multiple gameplay sessions under different music playback conditions [11], [18], [20], followed by post-gameplay questionnaires and optional physiological data collection, shown in Figure 3.

For music systems without real game integration, a purely listening test is often used instead of a game session [12], [29]. Where available, physiological and behavioural data, such as heart rate and skin conductance, are collected during gameplay [13]. This objective data complements the subjective feedback gathered through questionnaires. However, it is more difficult to obtain due to the experimental conditions. After each gameplay session, participants complete a questionnaire assessing player experience and music quality. The questionnaire includes multiple-choice questions, ranking/rating scales, and sometimes written feedback [18] to capture both quantitative and qualitative insights.

The collected data is then analysed using thematic analysis for qualitative feedback and statistical methods for quantitative data. This dual approach ensures a comprehensive evaluation of the procedural music generation systems. However, most studies investigated in this paper are primarily subjective, focusing on player experience and music quality; few papers have investigated the user experience of composers or sound designers working with the generation tools.

## 6 Challenges

By analysing research and application practices, we identify three key gaps. First, many academic systems are too experimental or computationally expensive for game developers to adopt. As shown in Figure 4, researchers focus on cutting-edge algorithms—such as reinforcement learning, genetic algorithms, and other advanced methods—that often require prerequisites for practical application, including high-quality datasets, real-time integration, and sophisticated fine-tuning. However, game developers often have to consider the overall allocation of resources in the game. Sound is often not the highest priority [31]. Therefore, they are less willing to experiment with risky or unconventional methods and instead prefer reliable, scalable, and extensible approaches, which has led to rule-based systems dominating the industry.

Second, the quality of many existing academic music systems is currently not up to industry standards. Based on the evaluation results of the academic research investigated in this paper, there is no strong evidence to suggest they provide a more enjoyable or immersive experience for players. A study of interviews with sound designers also confirms that music quality is a major contributor to the lack of acceptance of academic systems [8]. To simplify experimental design, many systems rely on high-level or single parameters—such as "emotion"—to control overall musical output [12], [13], [29], which significantly limits expressivity. In contrast, sound designers employ a wider range of game parameters to fine-tune various musical elements in real-world applications, yielding a much richer listening experience. Yet music quality depends on more than symbolic representation. Many sub-tasks of music generation—such as expressive rendering and modelling musical styles to match different gameplay scenarios—remain complex, unresolved challenges.

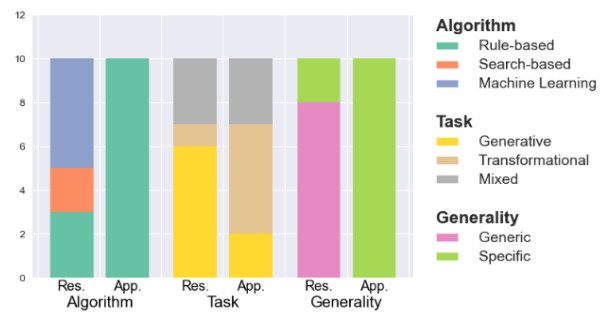

Figure 4. Comparison of Algorithm, Task and Generality in Research and Application

A third challenge is the absence of a standardised, user-friendly framework for integrating music systems into games[6]. Deploying these systems in a real game demands specialised expertise and considerable effort, since gameplay mechanics differ widely across styles and developers need solutions that plug directly into popular game engines. In practice, this often requires a dedicated team of audio engineers and access to proprietary assets, making it resource-intensive and sometimes legally complex. The fact that these music systems in academia are difficult to tailor to specific game contexts creates a gap that hampers the collection of in-game feedback and limits the real-world impact and evolution of academic work in this area.

## 7 Future Directions

Future research should focus on the following key areas to address the identified challenges.

A task-oriented approach that considers practical constraints such as available resources, deployment environments, and specific application requirements is advocated. When employing cutting-edge technologies like machine learning, focus on amplifying their benefits while thoughtfully addressing any limitations. By designing experiments with contextual awareness, researchers can create solutions that are both more extendable and industry-ready. At the same time, game developers should foster a spirit of experimentation with their music systems. Embracing a little risk can unlock richer, more immersive player experiences.

Beyond improving a music system's quality, it is even more important to establish evaluation criteria that accurately measure its contribution within the creative process. Such a framework must inclusively account for every participant involved in generating the music. While a long-term aim of Procedural Music Generation is to move beyond assisted creation toward fully autonomous composition, this shift must always serve to amplify human creativity. It is crucial to acknowledge the system's role as well as its users, ensuring we design tools that avoid stepping on the toes of our creatives [42], [43].

Integrating music into games involves a complex pipeline—DAWs, audio middleware, and the game engine—and many studios even build custom music and audio engines. To streamline procedural music system development, we need reusable, user-friendly research frameworks that plug seamlessly into popular game engines [16]. Studies like [13], [18] show how prototype games can serve as testbeds, giving researchers real-time player feedback to refine their systems and design more robust evaluations, despite the extra effort this requires. Ultimately, building a collaborative platform where researchers and developers share tools, datasets, and best practices will prove invaluable, accelerating innovation, fostering knowledge exchange, and bridging the gap between academic research and real-world game implementation.

## 8 Conclusion

This paper provides a comprehensive survey and analysis of PMG systems in both research and application domains. By employing a unified taxonomy, we compare these systems across various dimensions, identifying their development patterns, gaps, and challenges. Our analysis reveals that while advanced approaches and generative systems thrive in the research landscape, significant barriers must be conquered to achieve practical application, including deployment difficulties, music quality limitations, and a lack of seamless integration with games.

To address these challenges, we propose future directions that emphasise context-aware approaches tailored to specific tasks, music quality assessment considering human involvement in the creative process, and the development of enhanced research platforms and frameworks. We hope these efforts contribute to the long-term advancement of PMG, fostering both technological progress and practical application in the field.

**Appendix**

| Author/System | Year | Field | Generality | Task | Algorithm | Direction | Granularity | Grid | KS | Representation |
|---|---|---|---|---|---|---|---|---|---|---|
| **Spore [31]** | 2008 | Application | Specific | Mixed | Rule-based | Horizontal | Note | On | External | Symbolic |
| **Plans [21]** | 2012 | Research | Generic | Generative | GA | Horizontal | Note, Chord | On | External | Symbolic |
| **Proteus [34]** | 2013 | Application | Specific | Mixed | Rule-based | Mixed | Timbre | Off | External | Audio |
| **Sim Cell [32]** | 2014 | Application | Specific | Mixed | Rule-based | Mixed | Note | On | External | Symbolic |
| **Prechtl [13]** | 2016 | Research | Generic | Mixed | Markov chain | Horizontal | Chord, Tonality, Tempo, Velocity | On | External | Symbolic |
| **DOOM [35]** | 2016 | Application | Specific | Transformational | Rule-based | Horizontal | Timbre | On | External | Audio |
| **No Man's Sky [40]** | 2016 | Application | Specific | Transformational | Rule-based | Mixed | Phrase | On | External | Audio |
| **Genesis Noir [38]** | 2016 | Application | Specific | Transformational | Rule-based | Horizontal | Note | On | External | Symbolic |
| **Meta Compose [29]** | 2017 | Research | Generic | Mixed | Stochastic, GA, Rule-based | Mixed | Chord, Melody, Accompaniment | On | External | Symbolic |
| **Rez Infinite [37]** | 2017 | Application | Specific | Generative | Rule-based | Horizontal | Note | Off | External | Audio |
| **AMS [20]** | 2019 | Research | Generic | Transformational | GA, RNN, Graph Model | Horizontal | Note | On | Both | Symbolic |
| **Ape Out [36]** | 2019 | Application | Specific | Generative | Rule-based | Horizontal | Timbre | On | External | Audio |
| **Bardo Composer [12]** | 2020 | Research | Generic | Generative | Transformer, Beam Search | Horizontal | Note | On | Internal | Symbolic |
| **Amaral [27]** | 2022 | Research | Generic | Generative | Transformer | Mixed | Note, Timbre | On | Both | Symbolic |
| **PAMG [18]** | 2023 | Research | Specific | Mixed | Rule-based, Stochastic | Mixed | Note, Timbre | On | External | Symbolic |
| **Kopel [28]** | 2023 | Research | Generic | Generative | RNN | Horizontal | Note | On | Both | Symbolic |
| **Zumerle [11]** | 2023 | Research | Generic | Generative | CNN, Transformer | Horizontal | Note | On | Internal | Symbolic |
| **Tanabe [14]** | 2023 | Research | Specific | Generative | Rule-based | Horizontal | Pitch, Tempo, Velocity | On | External | Symbolic |
| **The Outlast trials [41]** | 2024 | Application | Specific | Transformational | Rule-based | Mixed | Phrase | On | External | Audio |
| **Ara: History Untold [39]** | 2024 | Application | Specific | Transformational | Rule-based | Mixed | Phrase | On | External | Audio |

Table 1. Key Procedural Music Systems in Research and Application